# LHC Detector Upgrades

Dan Green
dgreen@fnal.gov
## US CMS Dept., Fermilab

**talk presented at the**
LHC Symposium of May, 2003

The LHC detectors are well into their construction phase. The LHC schedule shows first beam to ATLAS and CMS in 2007. Because the LHC accelerator has begun to plan for a ten fold increase in LHC design luminosity (the SLHC or super LHC) it is none too soon to begin to think about the upgrades which will be required of the present LHC detectors. In particular, the tracking systems of ATLAS and CMS will need to be completely rebuilt. Given the time needed to do the R&D, make prototypes, and construct the new detectors and given the accelerator schedule for the SLHC, work needs to begin rather soon.

## Outline

There has already been a study in some depth about the physics reach and the needed upgrades to the LHC experiments, ATLAS and CMS, in the recent past [1]. Subsequently, there were presentations at the CERN ICFA meeting on the accelerator upgrades [2], the detector upgrades [3] and the consequent physics reach [4].

In this note only a brief overview of the detector needs is attempted. If there is a tenfold increase in luminosity (the Super LHC or SLHC). First an example of the increased physics reach using sequential Z bosons is given. The rapidity distribution of heavy states is also illustrated. Then a simple parameterization of the inclusive inelastic, or "minbias", events is presented.

This model is subsequently used to explore the impact of a tenfold increase in pileup on jet finding and reconstruction at the SLHC. It is also used to estimate the occupancy of tracker elements and the ionization radiation dose sustained by the tracker. Those estimates inform on the shape of possible tracker upgrades for the SLHC and the associated front-end electronics, which must also be upgraded.

The calorimetry of ATLAS and CMS must also be strengthened in order to work at the SLHC. The ATLAS liquid argon and the CMS crystal and scintillator calorimetry are briefly considered. A reduction in forward angular coverage to compensate for the increased radiation field is mentioned. For the muon system a similar reduction in angular coverage would maintain the remainder of the system in a state essentially the same as that for LHC operations. Finally, triggering and data acquisition issues are cursorily discussed.

**Physics Basics**

In order to get a feeling for the physics gains to be had with the SLHC, consider the production of a sequential Z of mass M. The production is assumed proceed by way of Drell-Yan annihilation of a quark-antiquark pair. The number, N, of detected Z depends on the luminosity, $\ell$, the electroweak fine structure constant, $\alpha_w$, the mass M, the C.M. energy, s, and the width in rapidity space over which the Z is produced, $\Delta y$ (which can be taken to be a constant). The number of produced Z' is:

$$N = \ell \{\pi^2 \alpha_W [xu(x) x\bar{u}(x)]_{x=M/\sqrt{s}}\} B(e^+e^- + \mu^+\mu^-)[\Delta y / 8M^2] \tag{1}$$

For quark distributions, $[xu(x)x\bar{u}(x)] = 0.36\sqrt{x}(1-x)^{11}$, if N = 100 is discovery level then M ~ 5.3 TeV is ~ the mass "reach" in 1 year at the SLHC where 4 TeV is the reach at the LHC (M = 4 -> 5.3 TeV). A plot of the maximum M as a function of luminosity for different C.M. energies is shown in Fig.1 where the Z' is assumed to have the same leptonic branching fraction as the Z.

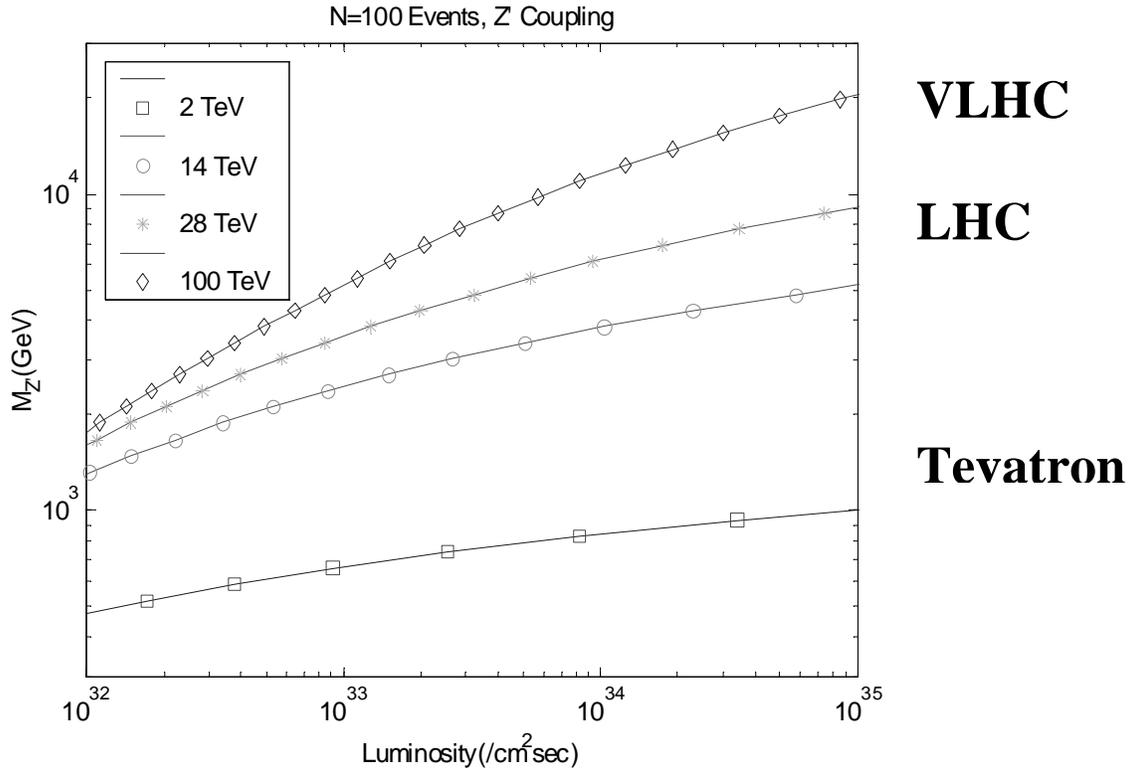

Figure 1: Required luminosity for discovery of a sequential Z boson for p-p colliders at four different C.M. energies

In general the mass reach is increased by ~ 30% for Z', heavy SUSY squarks or gluinos or extra dimension mass scales. A ~ 20% measurement of the HHH coupling is possible for Higgs masses < 200 GeV. However, to realize these improvements we need to maintain the capabilities of the LHC detectors.

It is important to show that all rapidities covered by the present LHC detectors are not equally populated in the study of high mass objects. For example, for a 1 TeV and 5 TeV Z' the distribution of the decay leptons in rapidity is given in Fig.2. Clearly, the leptons will be sharply limited to low |y| or large angles ("barrel"). Heavy states decay at wide angles defined purely by kinematics. Therefore, for these states we will concentrate preferentially on the wide-angle detectors.

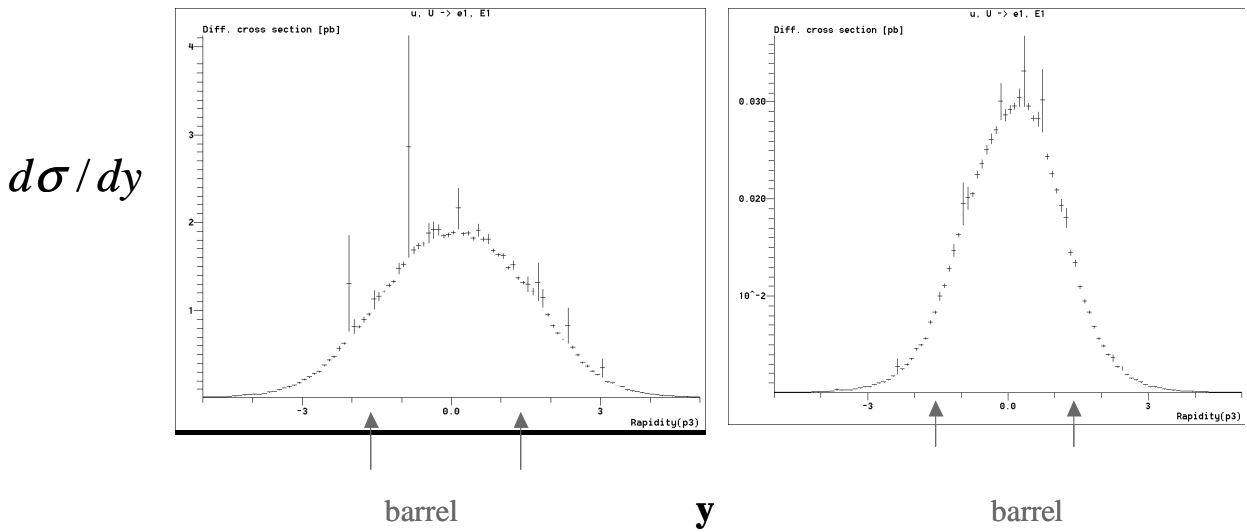

$d\sigma/dy$

barrel  y  barrel

Figure 2: Rapidity distribution of electrons from the decay of a sequential Z boson at the LHC for a 1 TeV and a 5 TeV boson. The arrows roughly indicate the barrel region.

## "Minbias" and Pileup

In order to get an approximate picture of the "pileup" of inclusive inelastic events at the LHC, a simplified model is attempted. A non-diffractive inelastic interaction at the LHC has several simple characteristics. The cross section is $\sigma_I$ which is ~ 50 mb. The interaction produces ~ equal numbers of $\pi^+, \pi^o, \pi^-$ which are distributed ~ uniformly in rapidity, y, with a "density" $\rho = (1/\sigma_I)d\sigma/dy$ ~ 9 pions per unit of y. The pions have a distribution in transverse momentum with a mean, $<p_T>$ ~ 0.6 GeV.

In this note it is assumed that the bunch spacing is reduced by a factor of two to 12.5 nsec. It is understood, that there may be accelerator issues to be confronted (e.g. electron cloud) which make this short bunch spacing unattainable. Nevertheless, under this assumption the number of interactions/crossing is only increased five-fold at the SLHC. The optimistic assumption is made that interactions are time resolvable between bunch crossings. In that case the pileup noise in the calorimeter is increased by 2.2 times. Again, if crossings are time resolvable in the tracker the occupancy of a detector element only increases fivefold. The tracker occupation can be estimated using a charged pion density of $\rho_c \sim 6$ charged pions per unit of rapidity, a luminosity $\ell = 10^{35}/cm^2 \cdot \sec$ and a bunch crossing time of 12.5 nsec. The expected parameters for operation at the LHC and SLHC are shown in Table 1.

Table 1: Parameters of the LHC and SLHC

|  | LHC | SLHC |
|---|---|---|
| $\sqrt{s}$ | 14 TeV | 14 TeV |
| L  /cm²sec | $10^{34}$ | $10^{35}$ |
|    fb⁻¹/yr | 100 | 1000 |
| Bunch spacing dt | 25 ns | 12.5 ns |
| N( interactions/x-ing) | ~ 12 | ~ 62 |
| $dN_{ch}/d\eta$ per x-ing | ~ 75 | ~ 375 |
| Tracker occupancy | $\equiv 1$ | 5 |
| Pile-up noise | $\equiv 1$ | ~2.2 |
| Dose central region | $\equiv 1$ | 10 |

**Jets at SLHC**

The pileup at the SLHC is expected to adversely effect the efficient and clean ( few fake jets) detection of low transverse momentum jets at ATLAS and CMS. In a cone of radius, R= 0.5 there are ~ 70 pions, or ~ 42 GeV of transverse momentum per crossing at the SLHC. This makes low Et jet triggering and reconstruction difficult.

There has been a heavy ion study [5] of Pb-Pb collisions where the density of produced charged pions is ~ 5000 per unit of rapidity. The charged particle density is similar to that for 833 p-p pileup events per bunch crossing or a luminosity ~ 13 times higher than the SLHC. Nevertheless, as seen in Fig. 3, the resolution of the jet energy is only degraded by 30% after an event-by-event procedure for pileup subtraction is applied. Indeed, the detection efficiency and purity of the jets is quite good for jet transverse energies above ~ 50 GeV.

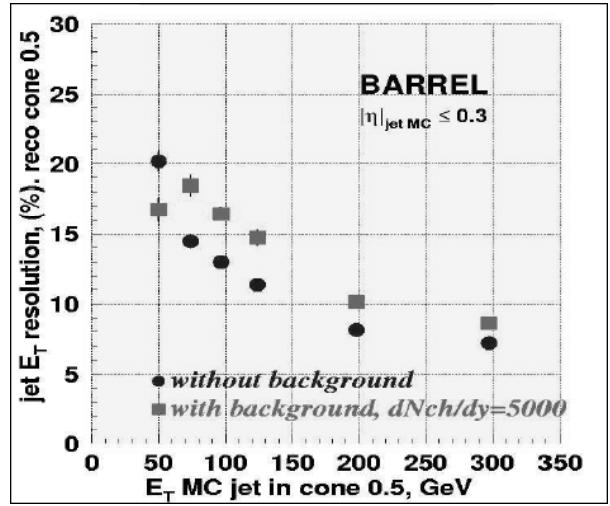

Figure 3: Energy resolution of jets in A-A collisions at the LHC. The charged particle density is ~ 5000 in this case, compared to ~ 6 in p-p collisions. The jet energy resolution is only degraded by about 30%.

However, the lower transverse energy jets that appear in vector boson fusion appear at small angles and are not without difficulties. These jets have $E_T \sim M_W/2$ and the "minbias" pileup within a cone of radius R = 0.5 and <y> ~ 3 is of a comparable magnitude in transverse energy (42 GeV).

A study of the impact of SLHC pileup on the probability of finding a fake "tag" jet is shown in Fig. 4. Clearly, there is a loss of a factor ~ 5 in fake rejection. To recover that loss one must use the energy flow inside a jet cone, specifically the existence of a high transverse momentum jet "core", to further reduce the fake jets due to pileup (which is~ uniform in R). The granularity of the LHC calorimeter, $\Delta\eta\Delta\phi = (0.087)^2$, means there are ~ 100 towers within a cone of radius R = 0.5, which allows for a detailed characterization of the energy flow within the jet.

It should also be possible to use the tracker to define "energy flow" inside the jet. Indeed, the tracker can also be used (energy flow mode) to subtract charged energy deposits from vertices within the crossing which are not of interest. That removal would further reduce the pileup energy by a factor 2/3.

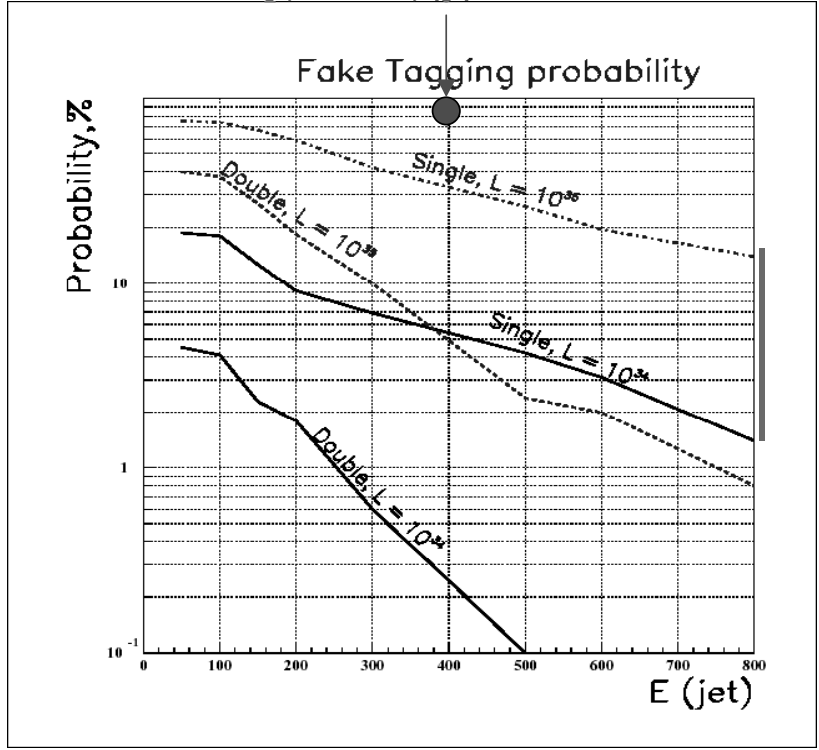

Figure 4: Plot of the probability for fake triggers on single and double "tag" jets at the LHC and the SLHC. Also indicated is the level of pileup in a cone of radius 0.5 and |y| = 3. Without careful mitigation, the single tag fakes rise dramatically in going to the SLHC.

Clearly, the tracker is crucial for much of the LHC physics [e.g. e, μ, jets (pileup, E flow), b tags]. However, as will be shown, the existing trackers will not be capable of utilizing the increased luminosity, as they will be near the end of their useful life. Therefore, it is necessary to completely rebuild the LHC tracking detectors. This will be the major element of the SLHC upgrade program. The other detectors do not require radical changes. Increasing occupancy of the detector elements, which causes pattern recognition difficulties, challenges the trackers. They also suffer from a ten-fold increase in radiation dose. These twin challenges define the requirements of the upgrades for the trackers.

## Occupancy and Radiation Dose

The occupancy, O, for a detector of area dA and sensitive time dt located at (r,z) is

$$O = \ell \sigma_I \rho_c (dAdt)/[2\pi r^2] \tag{2}$$

Clearly, Eq.2 holds for sensitive times greater than the bunch crossing time. For example, a silicon strip 10 cm x 100 μm with a 12.5 nsec sensitive time at r = 20 cm has a 1.5 % probability to be occupied by a charged track at the SLHC. For the higher luminosity burden of the SLHC one can decrease dA, or decrease dt (limit is the crossing time) or increase r ; get smaller, faster or further away.

The goal to is preserve the full tracker performance. If the silicon strips are pushed out to ~ 60 cm then the performance is ~ that of the LHC strips at 20 cm. This will require development. If one pushes the pixels out to 20 cm, the performance is roughly that of the LHC pixels at ~ 7 cm, again requiring development but nothing fundamentally new.

However, for r < 20 cm there is a pressing need for new technologies which will require basic research. Note that constant occupancy r preserves b tagging, but, with 12.5 nsec bunch crossing time, requires a five-fold pixel size reduction. Some possibilities are 3-d detectors with electrodes in bulk columns, diamond detectors (RD42) which are radiation hard, cryogenic detectors (RD39) which are fast and radiation hard, or monolithic detectors which should be cheaper and faster (reduced source capacity).

The tracker elements also have to stand a ten-fold increase in the radiation field at the SLHC. The ionizing dose, ID, due to charged particles is:

$$ID = \ell \sigma_I \rho_c \tau [dE/d(\rho'x)]_{mip} / [2\pi r^2] \qquad (3)$$

In terms of adjustable parameters, the dose depends only on luminosity, radius, and exposure time $\tau$. For example, at a radius of r = 20 cm, the SLHC dose is ~3 Mrad/yr. In this rough estimate one ignores heavily ionizing tracks, track curvature, magnetic capture ("loopers") and track interactions. In Fig.5 this is called the "naïve" expectation. Clearly, this estimate has the correct radial dependence, as verified by a detailed Monte Carlo program. The detailed estimates are higher, and the radiation doses are fierce. For example, the dose at 5 cm. is ~ 100 Mrad/yr. It is much to early to decide whether the full LHC tracking system capabilities can be maintained in the face of this hostile environment. An R&D program should be mounted very soon.

## Tracker Upgrade

It is conceptually useful to break up the tracker volume into radial regions. Roughly, with 10 fold increase in L, one needs a ~ three fold increase of radius to preserve an existing technology.

**Region 1 - r < 20cm:**
The occupancy requirement means that the pixel size should be a factor ~ five smaller than used today. For example, 125x125 $\mu m^2$ would perhaps become ~ 50 x 50 $\mu m^2$. This pixel shrinkage would also benefit b tagging. However, much R&D on fundamentally new pixel sensor technologies will be needed.

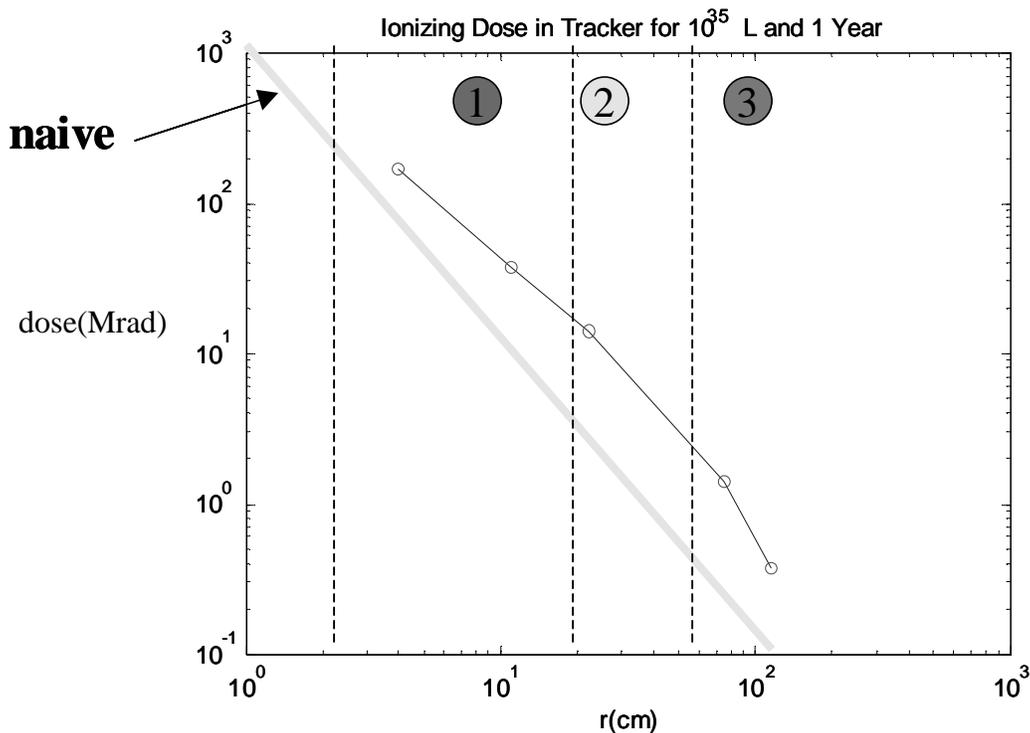

Figure 5: Ionizing radiation dose as a function of radial distance for the tracking system due to charged particle tracks at the SLHC. The naïve expectation is also shown.

**Region 2 - 20<r<60 cm:**
For this region one possible solution is to use pixel cell sizes ten times larger than current pixels but at a cost/channel ten times lower than current silicon microstrips. That would benefit the momentum resolution and improve pattern recognition. A less ambitious solution would be to upgrade the design of the innermost silicon microstrip layers of the current LHC detectors. R&D on thinner sensors to reduce interactions would be helpful.

**Region 3 - r > 60 cm:**
Clearly, the present designs can survive in this region. For the silicon strips pattern recognition argues to decrease the size of strips while maintaining the standard "radiation resistant" microstrip technology. R&D is still needed to follow commercial developments. For example, a feasibility study of processing detectors on 8" or 12" silicon wafers would be very useful.

**Engineering:**
The engineering should be put in place ab initio. There must be R&D on new materials, lightweight and stable structures, cooling, alignment, cryogenic operations, installation and maintenance. For example, an expected tracker activation of ~ 250 mSv/hr has serious implications for access and maintenance.

**Micro-electronics:**

The front-end electronics for the tracker is closely coupled to the sensors. Indeed, it may be a monolithic three-dimensional device with integrated electronics readout. It has been true for some time, see Fig.6, that line-widths decrease by a factor of two every five years. For example, deep sub-micron (DSM = 0.25 µm) is radiation hard and widely used at the LHC. Today 0.13 μm is commercially available (Fig.6). Meanwhile, in the research lab 0.04 μm, e.g. extreme UV lithography, is in existence. Therefore, one can confidently predict that the trend will continue for a decade, which allows an extrapolation to the time when the SLHC upgrade purchases are scheduled.

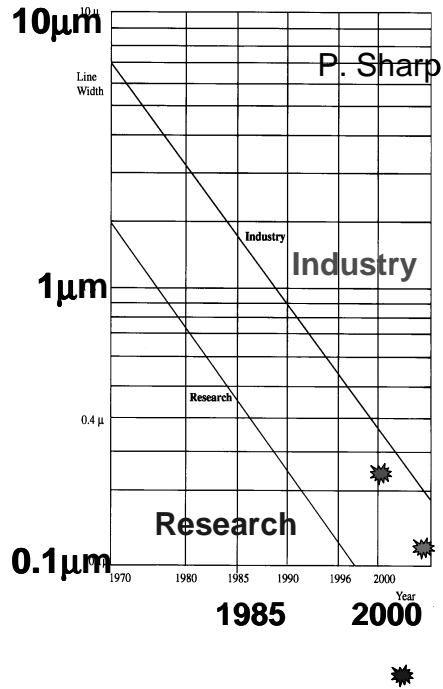

Figure 6: Feature size available in electronics both in the R&D phase and in the commercially available processes as a function of calendar year. Recent developments are shown as stars in the plot.

A possible electronics R&D plan is to characterize emerging technologies. In particular, more radiation tolerance is going to be required. This is not just an issue of dose but also of single event effects (upset and latch up) with the smaller feature sizes. In addition, the commercial arena shows rapid growth in advanced high bandwidth data link technologies. In this case one wants to address electronics system issues from the start.

**Timescale:**
The necessary preparation will obviously need ~ 4 years of R&D and prototyping. The full upgrade will need 8-10 years from the first launch of R&D, assuming ~ 4-6 years of construction. Given that the LHC accelerator is discussing the time period ~ (2012-2014) for the SLHC, it is required that ATLAS and CMS begin the R&D program rather soon.

## Calorimeter Dose and Operation

The calorimeters operating at the LHC are at larger radii than the trackers. However, the full energy of the secondary tracks is absorbed, exposing the calorimetry to larger energy deposits per track than the trackers. The showering dose (SD) in the electromagnetic calorimeter (ECAL) is ~ due to photon showers and is:

$$SD = \ell \sigma_I \rho_o \tau [dE/d(\rho' x)]_{mip} [<p_T>/\sin\theta E_c]/[2\pi r^2]$$
$$= (ID/2)[<p_T>/\sin\theta E_c] \quad (4)$$

Basically, the dose is the ionizing dose (ID) enhanced by the number of particles at shower maximum, which is roughly the photon energy divided by the critical energy, $E_c$, in the material. In the barrel, SD is ~ $\ell/[r^2 \sin\theta]$ which has only a weak angular dependence. In the endcap however, SD ~ $\ell/[z^2 \theta^3] \sim (\ell/z^2)e^{3\eta}$ which has a very strong angular dependence. For example, at r = 1.2 m, for Pb with $E_c$ = 7.4 MeV, the dose is estimated to be 3.3 Mrad/yr at y=0, and 7.8 Mrad/yr at |y|=1.5. The results of a detailed Monte Carlo calculation and these "naïve" estimates are shown in Fig.7. The basic angular dependence of the "naïve" estimates is confirmed.

The dose ratio of the hadron calorimeter (HCAL) to the ECAL is due to the different energy thresholds for shower multiplication and is ~ $E_{th}(\pi + p \to \pi + \pi + p)/E_c$ where the threshold for pion production is $E_{th}$. This rough estimate for the ratio is also observed in Fig.7.

The barrel doses are not a problem for the LHC calorimeters. However, for the endcaps a technology change may be needed for 2 < |y| < 3 for the CMS HCAL. One possibility is to switch from scintillator (Fig.8) to the quartz fibers which are already in use in the forward calorimetry, 3<|y|<5, using a technology that works at the LHC up to |y|~ 5 and migrating it to |y|~3 guided by the factor $e^{3\eta}$ in the dose. The ECAL of CMS has APD leakage current noise issues in the endcap as well as radiation damage questions, which will require development.

The ATLAS LA calorimetry has space charge and current draw issues, which become worrisome for the small angle regions of the endcap ECAL. The region of critical energy density deposit is shown in Fig.9. These questions also require development rather than fundamentally new detectors. Possible ATLAS solutions are an alternative cryogenic liquid or a cold dense gas.

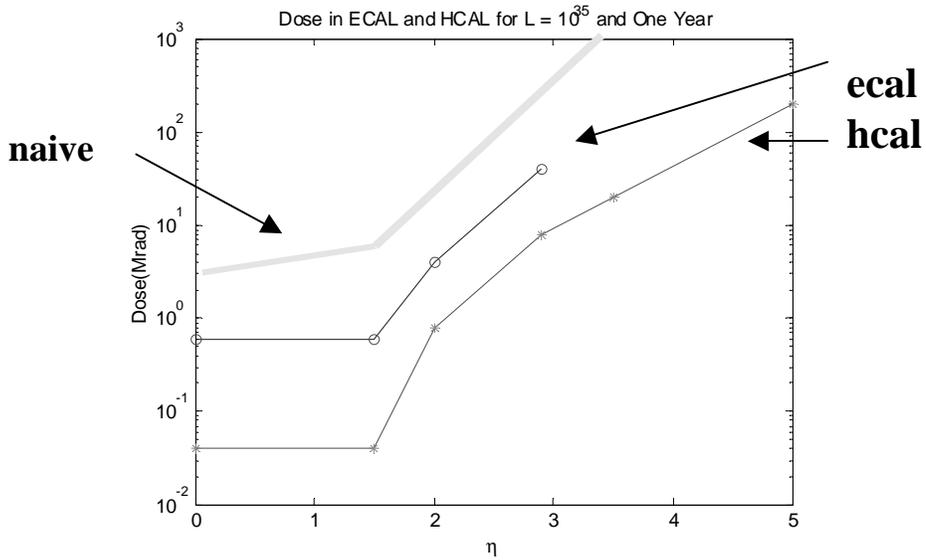

Figure 7: Shower dose in the CMS calorimetry as a function of pseudorapidity for both the electromagnetic and hadronic compartments. The naïve estimate is also shown for comparison.

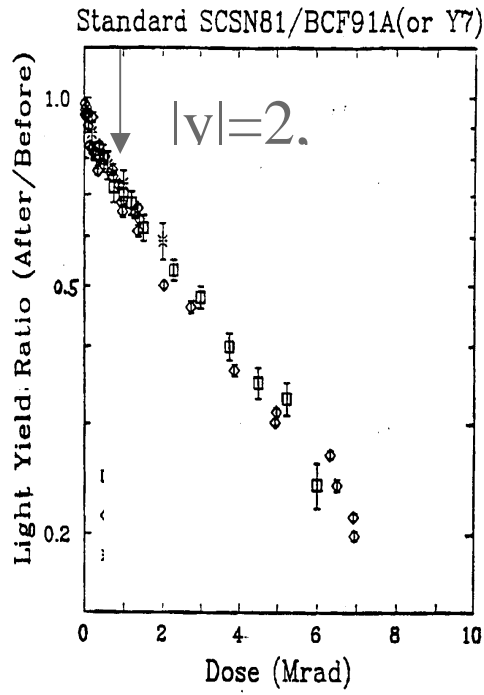

Figure 8: Loss of light in plastic scintillator as a function of radiation dose. The arrow indicates the dose in the hadron calorimeter at $|y|=2$ for one year of operation at the SLHC.

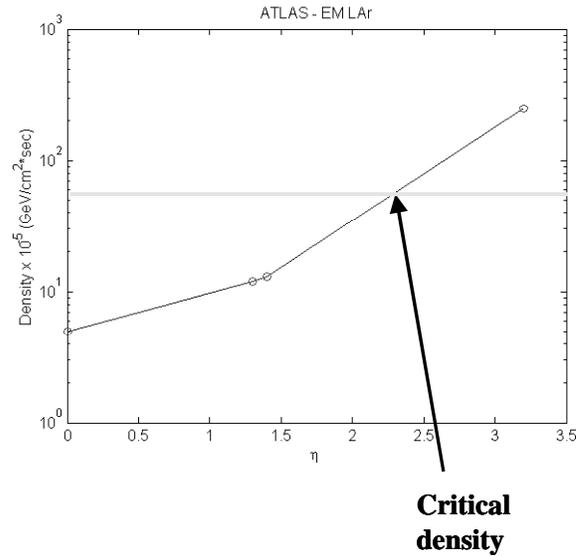

**Critical density**

Figure 9: Ionization density in the ATLAS liquid argon calorimeter as a function of pseudorapidity of the electromagnetic compartment.

It is clear, Eq.4, that the worst difficulties of radiation dose occur at the smallest angles. One simple possibility is to reduce the forward coverage to compensate for the ten-fold SLHC luminosity increase. For the case of vector boson fusion production, Fig.10, this reduction is not too damaging to "tag jet" efficiency. A reduction to y < 4.2 naively keeps the dose constant. As seen in Fig.10, the peak "tag" rate occurs at |y|=3. If deemed necessary, one could replace the CMS quartz fibers with high-pressure gas. This is an area where development is needed.

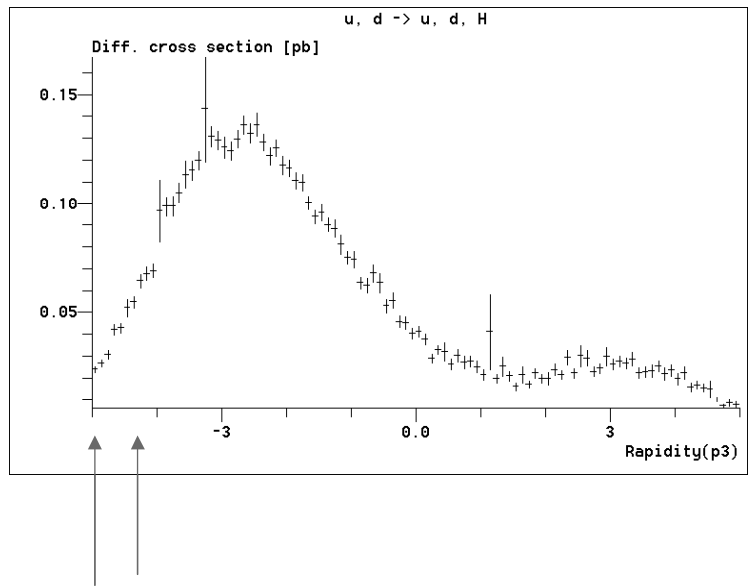

Figure 10: Rapidity distribution of the "tag" quarks in vector boson fusion production of a light Higgs boson. The arrows show the CMS angular coverage and the coverage that would yield the same radiation dose as |y|=5 at the SLHC.

## Muon Systems

Typically, for the LHC muon detectors there is factor ~ five in "headroom" at design L. For the SLHC there are several possibilities. One is just to maintain the angular coverage and develop more radiation resistant technologies. An alternative possibility is to acknowledge that one is probably studying massive states at the SLHC (Fig.2) and to consider reduced angular coverage, as was done for the calorimetry.

In addition, the reduced coverage would allow for alternative shielding schemes. With this added shielding, the dose rates in the muon system (and other forward detectors) can be kept constant if the angular coverage goes from |y|<2.4 to |y|<2. A detailed comparison of the shielding plans is shown in Fig.11. It appears that the neutron background can be controlled with this minor reduction in angular coverage. More detailed physics studies will be needed to assess the real impact on the SLHC physics program.

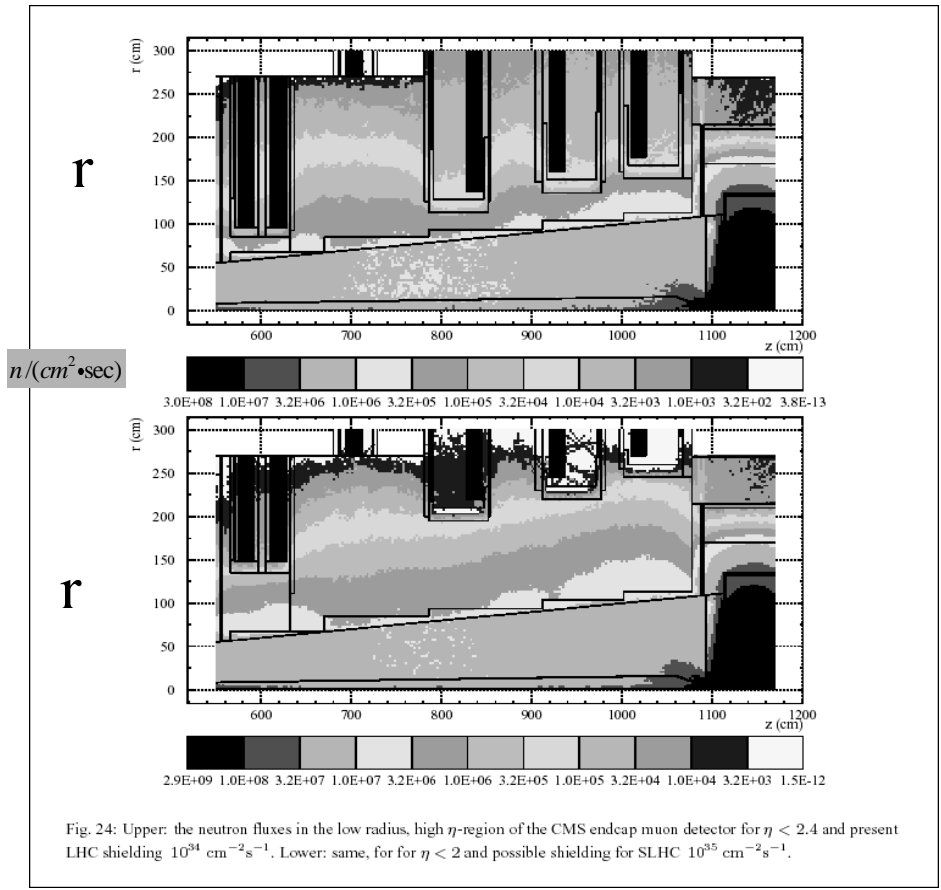

Fig. 24: Upper: the neutron fluxes in the low radius, high $\eta$-region of the CMS endcap muon detector for $\eta < 2.4$ and present LHC shielding $10^{34}$ cm$^{-2}$s$^{-1}$. Lower: same, for for $\eta < 2$ and possible shielding for SLHC $10^{35}$ cm$^{-2}$s$^{-1}$.

Figure 11: Neutron fluences at CMS for the LHC and the SLHC configurations of shielding

## Trigger and Data Acquisition

The trigger and data acquisition (DAQ) subsystems will be largely driven by commercial developments. Assuming that the LHC initial physics program is successful, one can simply raise the trigger thresholds. That just mirrors the concept that SLHC is moving forward to study heavier mass states.

If the bunch crossing time can be reduced it is advantageous to rebuild the trigger system to run at 80 MHz. For triggering, one then tries to utilize those detectors that are fast enough to give a bunch crossing identification within 12.5 nsec (e.g. Calorimetry, Tracking).

In general, developments in front end electronics will allow more intelligence to reside before the L1 trigger. These advances should be carefully tracked and then exploited. The DAQ must either design for the increased event size (pileup) with reduced first level (L1) trigger rate or accept data compression and design a benign method to reduce the data volume. For the DAQ system the rapid advances in commercial technology argue to carefully track the evolution of communication technologies, e.g. 10 Gb/sec Ethernet.

## Summary

The LHC physics reach will be substantially increased by the higher luminosity available at the SLHC. In order to realize that improvement, the LHC detectors must be adapted at the SLHC to preserve their LHC performance. The trackers must be rebuilt with a fundamentally new technology at radii below 20 cm. That upgrade will be the major item in the SLHC upgrade program. The calorimeters, muon systems, triggers and DAQ will need development but not wholesale replacement. The upgrades are likely to take ~ (6-10) years. The SLHC accelerator is thought to be ready ~ (2012, 2014). Given that the development time for the present LHC detectors was at least a decade, the time to start is obviously "now", and the people to do the job are those who did it for the present detectors. Since the LHC detectors are not yet commissioned, new people are sorely needed, but mixed with the present LHC developers because the upgrades are partially an R&D effort but constrained by the necessity to integrate into the existing ATLAS and CMS detectors.